\documentclass[aps,pre,floatfix,superscriptaddress]{revtex4}
\usepackage{graphicx}
\usepackage{bm}
\bibstyle{apsrev.bib}

\begin{document}

\title{Exact enumeration of self-avoiding walks}
\author{R. D. Schram}
\affiliation{Institute for Theoretical Physics, Utrecht University,
P.O. Box 80195, 3508 TD  Utrecht, The Netherlands}
\affiliation{Mathematical Institute, Utrecht University,
P.O. Box 80010, 3508 TA Utrecht, The Netherlands}
\author{G. T. Barkema}
\affiliation{Institute for Theoretical Physics, Utrecht University,
P.O. Box 80195, 3508 TD  Utrecht, The Netherlands}
\author{R. H. Bisseling}
\affiliation{Mathematical Institute, Utrecht University,
P.O. Box 80010, 3508 TA Utrecht, The Netherlands}

\date{\today}

\begin{abstract}
A prototypical problem on which techniques for exact enumeration are tested
and compared is the enumeration of self-avoiding walks.  Here, we show an
advance in the methodology of enumeration, making the process thousands
or millions of times faster. This allowed us to enumerate self-avoiding walks
on the simple cubic lattice up to a length of 36 steps.
\end{abstract}


\maketitle

\section{Introduction}

According to renormalization group theory, the scaling properties of
critical systems are insensitive to microscopic details and are governed
by a small set of universal exponents.  Polymers can be considered
as critical systems in the limit where their length $N$ (the number
of chained monomers) grows~\cite{degennes}. For instance, the free energy $F_N$ of an
isolated polymer in a swollen phase behaves asymptotically as $\exp(-F_N) \equiv Z_N \approx A \mu^N N^\theta$.
Here, the connectivity constant $\mu$ and the amplitude $A$ are
non-universal (model-dependent) quantities. The exponent $\theta$,
however, characterizing the leading correction to the scaling behavior,
is believed to be universal; it is related to the entropic exponent
$\gamma_s = \theta + 1$.
The average squared distance between the end points of such polymers
scales as $N^{2\nu}$, where $\nu \approx 0.588$ in three dimensions
is also a universal critical exponent.

Universal exponents such as $\theta$ and $\nu$ can be measured most accurately in
computer simulations of the most rudimentary models in the universality
class of swollen polymers, which arguably is that of self-avoiding walks
(SAWs) on a lattice. Estimates of these exponents can be obtained by
counting the number $Z_N$ of SAWs of all lengths up to $N_{\max}$, and calculating
the sum $P_N$
of their squared end-to-end extensions, which scales as $P_N \sim Z_N N^{2\nu}$.
The exponents can then be obtained
from
\begin{equation}
\label{eqn:theta}
\theta = \frac{N^2-4}{4} \left[ \log \frac{Z_N^2}{Z_{N+2} Z_{N-2}} \right]
\end{equation}
and
\begin{equation}
\label{eqn:nu}
\nu = \frac{N-1}{4} \left[ \log \frac{P_{N+1}}{Z_{N+1}} - \log \frac{P_{N-1}}{Z_{N-1}} \right],
\end{equation}
respectively, in the limit of increasing $N$.
In Eq.~(\ref{eqn:theta}),
the values of $N$ are taken a distance two apart, 
so that the formula involves either only even $N$, or only odd $N$; 
this is more accurate than mixing even and odd values.
Similar considerations lead to Eq.~(\ref{eqn:nu}).
The accuracy of the estimates improves significantly with increasing
$N_{\max}$, but unfortunately at the expense of an exponentially growing
number of walks. In two dimensions, various algorithmic improvements
have allowed for the enumeration of all SAWs up to $N_{\max}=71$ steps~\cite{jensen}, but
these methods cannot be used effectively in three dimensions, which is
the most relevant dimensionality for practical purposes. Hence, to date,
the enumeration of three-dimensional SAWs stops at $N_{\max}=30$ steps~\cite{clisby}. 

Counting SAWs has a long history, see e.g. \cite{madras96}.
In a paper by Orr~\cite{orr47} from 1947, $Z_N$ was given for all $N$ up to 
$N_{\max}=6$; these values were calculated by hand.
In 1959, Fisher and Sykes~\cite{fisher59} enumerated all SAWs in 3D 
up to $N_{\max}=9$ using a computer.
More recently, in 1987 Guttmann~\cite{guttmann87} enumerated longer SAWs up to $N_{\max}=20$,
and extended this by one step in 1989~\cite{guttmann89}.
In 1992, MacDonald et al.~\cite{macdonald92} reached $N_{\max}=23$,
and in 2000 MacDonald et al.~\cite{macdonald00} reached $N_{\max}=26$.

Here, we present the new length-doubling method which allowed
us to reach $N_{\max}=36$ using 50,000 hours of computing time,
a result that would have taken roughly fifty million hours with
traditional methods, or alternatively we would have to wait another
20 years by Moore's law (which states that the number of transistors on a computer chip doubles every two years) before we could undertake the
computation.

\section{Length-doubling method}

In the length-doubling method, we determine for each non-empty subset
$S$ of lattice sites the number $Z_N(S)$ of SAWs with length $N$ and
originating in the origin, that visit the complete subset.  Let $|S|$
denote the number of sites in $S$. The number $Z_{2N}$ of SAWs
of length $2N$ can then be obtained by the length-doubling formula

\begin{equation}
\label{eqn:doubling}
Z_{2N} = Z_N^2 + \sum_{S \neq \emptyset} (-1)^{|S|} Z_N^2(S) .
\end{equation}

This equation can be understood as follows.
Let $N \geq 1$ be fixed. Let $A_i$ be the set of pairs $(v,w)$ of SAWs
of length $N$ that both pass through lattice point $i$. 
Here, a walk $v$ starts in 0, and then passes through $v_1, \ldots, v_N$.
Since the distance reached
from the origin is at most $N$, there exist only finitely many non-empty sets 
$A_i$. Then, the total number of 
SAWs of length $2N$ equals
\begin{equation}
\label{eqn:z2na}
Z_{2N} = Z_N^2 - \left| \bigcup_i A_i \right|,
\end{equation}
because every pair $(v,w)$ of the $Z_N^2$ possible pairs 
can be used to construct a SAW of length $2N$,
except if $v$ and $w$ intersect in a lattice point $i$.
The resulting walk
\begin{equation}
\label{eqn:vw}
(v,w) \equiv (v_{N-1}- v_N, \ldots,  v_1-v_N, -v_N,w_1-v_N, \ldots, w_N - v_N)
\end{equation}
of length $2N$ is obtained by connecting the two walks in 0 and translating the result over a distance
$-v_N$. The new starting point 0 is then the translated old end point of $v$
and the new end point is the translated old end point of $w$.
Note that from a SAW of length $2N$ we can also create a 
non-intersecting pair $(v,w)$ by using (\ref{eqn:vw}), so that indeed we have a bijection
between such pairs and SAWs of length $2N$.

\begin{figure}
\includegraphics[width=14cm]{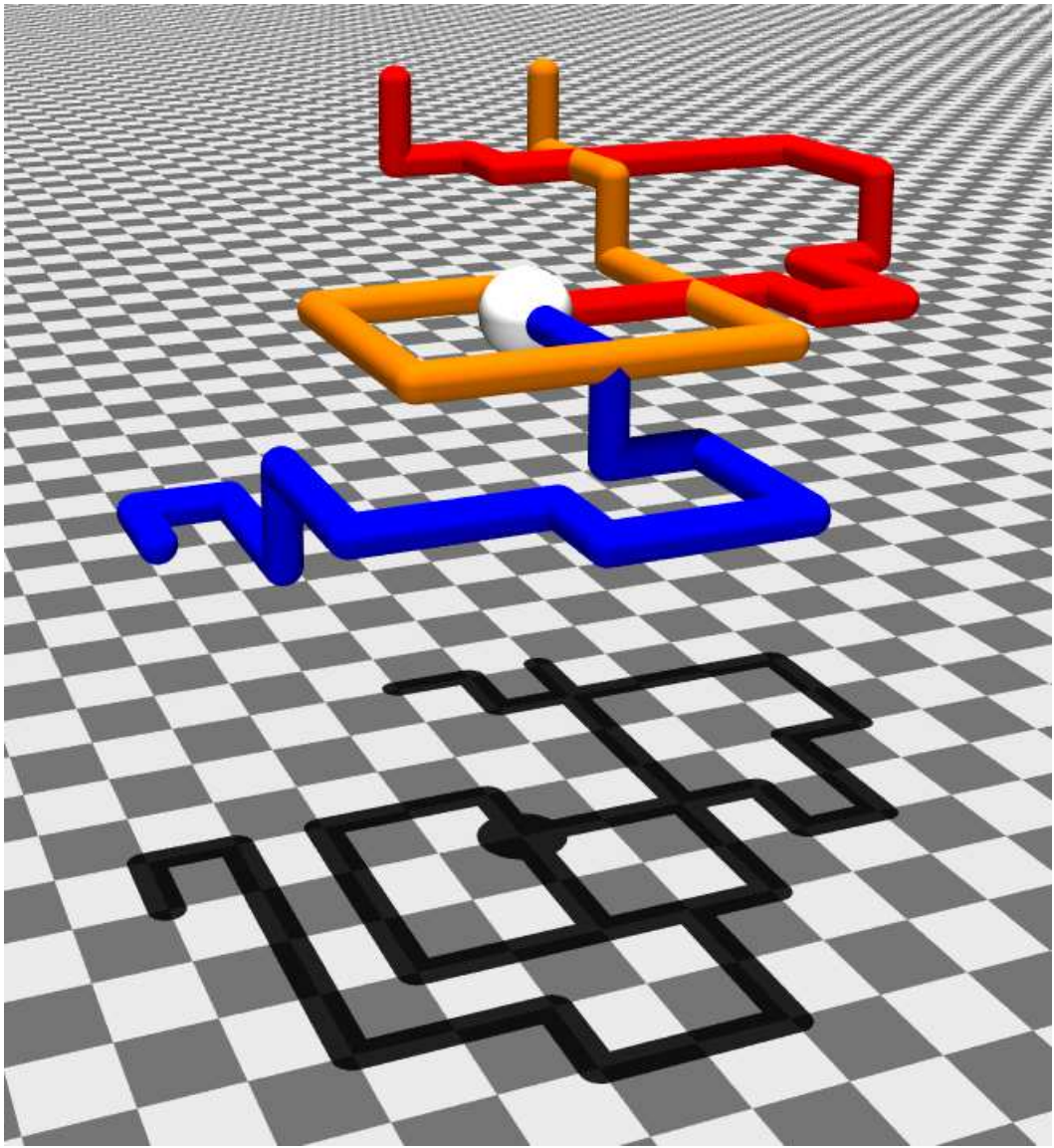}
\caption{
Illustration of the length-doubling algorithm, using a small
subset of three walks of length $N=18$. Ignoring intersections, there
are $Z=3$ candidates for SAWs of length 36: the blue-red, blue-orange,
and red-orange combinations. Ignoring double counting, $Z$ should be
reduced by 3 because of the intersections $a=(2,3,1)$, $b=(2,0,0)$, and $c=
(0,-2,0)$. Correcting for double counting because of the pair of sites
$S= \{a,b\}$,  the number of self-avoiding combinations is thus $3-3+1=1$. Indeed,
only the red-blue combination is self-avoiding. Using a computer,
we applied this approach to combinations of \emph{all} walks of length $N=18$.}
\label{fig:doubling}
\end{figure}

The inclusion-exclusion principle from combinatorics, 
see for instance \cite[Chapter 10]{vanlint92}, states that
\begin{equation}
\label{eqn:inclexcl}
\left| \bigcup_{i=1}^n A_i \right| = \sum_i |A_i| - \sum_{i<j} |A_i \cap A_j| + 
\sum_{i<j<k} |A_i \cap A_j \cap A_k|
+ \cdots +
(-1)^{n+1} |A_1 \cap A_2 \cdots \cap A_n|,
\end{equation}
for the union of $n$ sets $A_i$.
We can apply this principle, noting that for a non-empty set $S = \{ i_1, \dots , i_r\}$
the intersection $A_{i_1} \cap \cdots \cap  A_{i_r}$ has $Z_N^2(S)$ elements,
where $Z_N(S)$ is defined as the number of SAWs of length $N$
that pass through all the sites of $S$. The sign of the term corresponding to the set $S$
in the expansion~(\ref{eqn:inclexcl}) is $(-1)^{r+1}$, where $r = |S|$.
Substituting this in Eq.~(\ref{eqn:inclexcl}) and combining with Eq.~(\ref{eqn:z2na})
yields the length-doubling formula Eq.~(\ref{eqn:doubling}).
The length-doubling method is illustrated by Fig.~\ref{fig:doubling}.

\section{Application of the length-doubling formula}

The usefulness of this formula lies in the fact that the numbers $Z_N(S)$ can
be obtained relatively efficiently:
\begin{itemize}
\item Each SAW of length $N$ is generated.
\item For each SAW, each of the $2^N$ subsets $S$ of lattice sites
is generated, and the counter for each specific subset is incremented.
Multiple counters for the same subset $S$ must be avoided; this can
be achieved by sorting the sites within each subset in an unambiguous way.
\item As the last step, the squares of these counters are summed, with a positive
and negative sign for subsets with an even and odd number of sites, respectively,
as in Eq.~(\ref{eqn:doubling}).
\end{itemize}

With $Z_N$ walks of length $N$, each visiting $2^N$ subsets of sites,
the computational complexity is $\mathcal{O}(2^N Z_N) \sim (2\mu)^N$ times
some polynomial in $N$ which depends on implementation details. This
compares favorably to generating all $Z_{2N} \sim \mu^{2N}$ walks of
length $2N$, as $\mu \approx 4.684$ on the simple cubic lattice.

A practical problem which is encountered already at relatively low $N$,
is the memory requirement for storing the counters for all subsets. An
efficient data structure to store these is based on a tree structure. The
occurrence of a subset $\{a,b,c,d,e\}$, in which $a,b,c,d,$ and $e$ are site
numbers ordered such that $a<b<c<d<e$, is stored in the path $a \rightarrow
b\rightarrow c\rightarrow d\rightarrow e$,
where $a$ is directly connected to the root of the tree
and $e$ is a leaf.

We added two further refinements to the method sketched above. First,
we exploit symmetry.  Two subsets $S_1$ and $S_2$ which are related by
symmetry will end up with the same counter.  One can therefore safely
keep track of the counter belonging to only one subset $S$ out of
each group of symmetry-related subsets. This reduces the memory
requirement by a factor close to 48 (slightly less because of subsets
with an inherent symmetry); in practice, the computational effort
goes down by a similar factor.

The second refinement is tree splitting. Rather than computing the
full tree, we split the tree into non-overlapping subtrees, using
for instance as a criterion the value of the site with the highest
number. Another criterion is the subset size $|S|$.
This splits up the summation in Eq.~(\ref{eqn:doubling}) into
independent sums, which can be computed in parallel.

With the length-doubling method, it is also possible to compute the
squared end-to-end distance, summed over all SAW configurations.
The squared end-to-end distance for walks of length $N$
is defined by 
\begin{equation}
\label{eqn:PN}
P_N = \sum_w ||w_N||^2,
\end{equation}
where the sum is taken over all the SAWs of length $N$,
and $||w_N||$ is the Euclidean distance of the end point $w_N$ of walk $w$
from the origin.

The length-doubling formula for the squared end-to-end distance
then becomes
\begin{equation}
\label{eq:distdoubling}
P_{2N} = 2Z_N P_N + 2\sum_{S \neq \emptyset} (-1)^{|S|} 
\left( Z_N(S)P_N(S) -||E_N(S)||^2 \right) .
\end{equation}
Here, $P_N(S)$ is the total squared end-to-end distance 
for all walks of length $N$ that pass through the complete set $S$,
and the extension $E_N(S)$ is defined as the sum of $w_N$ for all such walks $w$.
This formula can be understood again by using the inclusion-exclusion
principle, but now generalised to add (squared) distances for sets $A_i$
instead of just counting numbers of elements.
The first term of the right-hand side of Eq.~(\ref{eq:distdoubling})
is obtained by computing 
\begin{eqnarray}
\label{eqn:ZNPN}
\sum_{(v,w)} ||w_N-v_N||^2 & = &
\sum_{(v,w)} (||w_N||^2+||v_N||^2 -2 v_N \cdot w_N) \nonumber \\
& = & Z_N \sum_w ||w_N||^2 + Z_N \sum_v ||v_N||^2 
    -2 (\sum_v v_N) \cdot (\sum_w w_N) \nonumber \\
& = & 2Z_N P_N,
\end{eqnarray}
where the inner product vanishes because of the symmetry
between $v$ and $-v$.
For walks passing through $S$ a similar derivation holds,
but now the inner product does not vanish,
and instead gives rise to the term $||E_N(S)||^2$.
Computing $P_{2N}$ by this formula requires additional counters for each subset $S$,
namely for the total extension in the $x$-, $y$- and $z$-directions,
as well as for the total squared extension $P_N(S)$.

\section{Results}

With length-doubling, we obtained $Z_N$ up to $Z_{36}= 2~941~370~856~334~701~726~560~670$,
with a squared end-to-end extension of $P_{36}= 230~547~785~968~352~575~619~933~376$.
All values of $Z_N$ and $P_N$ for $N \leq 36 $ are given in Table~\ref{tab:data}.
In a first analysis of the results, we fitted the data for $N=18,\ldots,36$ with the functions
\begin{eqnarray}
\tilde{Z}_N & \approx & A_1 \mu_1^N N^{\theta_1} \left( 1 + c_1 N^{-\Delta_1} + k_1 (-1)^N N^{-\alpha_1} \right), \nonumber \\
\tilde{P}_N & \approx & A_2 \mu_2^N N^{\theta_2} \left( 1 + c_2 N^{-\Delta_2} + k_2 \frac{1+(-1)^N}{2} N^{-\alpha_2} \right),
\label{eq:fits}
\end{eqnarray}
by minimizing $\epsilon_1=\sum_{i=18}^{36} \left( \log \tilde{Z}_i-\log Z_i \right)^2$
and $\epsilon_2=\sum_{i=18}^{36} \left( \log \tilde{P}_i -\log P_i \right)^2$
with respect to the fitting parameters.
The different treatment of the oscillatory corrections, $(-1)^N$ vs. $(1+(-1)^N)/2$,
gave us a slightly better fit, but we do not attribute much physical significance to this.
The resulting fit
values are presented in Table~\ref{tab:fits}. Note that the resulting values
for $\epsilon_1$ and $\epsilon_2$ are small, indicating that the deviations between
the exact number $\log Z_i$ and the approximation $\log \tilde{Z}_i $ only show up
in the ninth or tenth significant digit.

From a physics perspective, the most interesting observables are the
growth exponent $\nu$, and the entropic exponent $\gamma_s$. The fitting procedure
yields $\nu=(\theta_2-\theta_1)/2=0.593$, which is to be compared with the
Monte Carlo result $\nu=0.587597(7)$ as obtained by Clisby~\cite{pivot}.
The fitting procedure also yields $\gamma_s=\theta_1+1=1.1597$, which is
not too far from the literature value $\gamma_s=1.1573(2)$ as obtained
by Hsu et al.~\cite{hsu} using the pruned-enriched Rosenbluth method.
Deviations between our results and those in the current literature are
likely related to the fact that we observe finite-size corrections with
correction exponents $\Delta_1\approx 1.4$ and $\Delta_2\approx 0.26$,
while the literature values for $\nu$ and $\gamma_s$ are obtained under
the assumption of a shared correction exponent of $\Delta \approx 0.46$.
A more refined analysis of our enumeration data, along the lines
of Ref.~\cite{clisby}, is required to investigate the nature of the
finite-size corrections.
Such an analysis will probably also yield slightly different values for $\nu$ and 
$\gamma_s$.

In the near future, we will apply our new approach for exact enumeration
to other lattices such as face-centered-cubic and body-centered-cubic, and
generalize it to various other models in polymer physics, such as confined
and branched polymers, and to various other models in statistical physics.

\section{Acknowledgement}

Computations were carried out on the Huygens supercomputer at SARA in Amsterdam.
We thank the Dutch National Computing Facilities Foundation (NCF)
for providing us with computer resources under the project SH-174-10.

\begin{table}
\caption{
\label{tab:data}
Enumeration results on the number of three-dimensional
self-avoiding walks $Z_N$ and the sum of their squared end-to-end distances $P_N$.}
\begin{tabular}{lrr}
$N$ & $Z_N$ & $P_N$ \\
\hline
1&6\,&\,6\,\\
2&30\,&\,72\,\\
3&150\,&\,582\,\\
4&726\,&\,4\,032\,\\
5&3\,534&25\,566\,\\
6&16\,926&153\,528\,\\
7&81\,390&886\,926\,\\
8&387\,966&4\,983\,456\,\\
9&1\,853\,886&27\,401\,502\,\\
10&8\,809\,878&148\,157\,880\,\\
11&41\,934\,150&790\,096\,950\,\\
12&198\,842\,742&4\,166\,321\,184\,\\
13&943\,974\,510&21\,760\,624\,254\,\\
14&4\,468\,911\,678&112\,743\,796\,632\,\\
15&21\,175\,146\,054&580\,052\,260\,230\,\\
16&100\,121\,875\,974&2\,966\,294\,589\,312\,\\
17&473\,730\,252\,102&15\,087\,996\,161\,382\,\\
18&2\,237\,723\,684\,094&76\,384\,144\,381\,272\,\\
19&10\,576\,033\,219\,614&385\,066\,579\,325\,550\,\\
20&49\,917\,327\,838\,734&1\,933\,885\,653\,380\,544\,\\
21&235\,710\,090\,502\,158&9\,679\,153\,967\,272\,734\,\\
22&1\,111\,781\,983\,442\,406&48\,295\,148\,145\,655\,224\,\\
23&5\,245\,988\,215\,191\,414&240\,292\,643\,254\,616\,694\,\\
24&24\,730\,180\,885\,580\,790&1\,192\,\,504\,522\,283\,625\,600\,\\
25&116\,618\,841\,700\,433\,358&5\,904\,\,015\,201\,226\,909\,614\,\\
26&549\,493\,796\,867\,100\,942&29\,166\,\,829\,902\,019\,914\,840\,\\
27&2\,589\,874\,864\,863\,200\,574&143\,797\,\,743\,705\,453\,990\,030\,\\
28&12\,198\,184\,788\,179\,866\,902&707\,626\,\,784\,073\,985\,438\,752\,\\
29&57\,466\,913\,094\,951\,837\,030&3\,476\,154\,\,136\,334\,368\,955\,958\,\\
30&270\,569\,905\,525\,454\,674\,614&17\,048\,697\,\,241\,184\,582\,716\,248\,\\
31&1\,274\,191\,064\,726\,416\,905\,966&83\,487\,969\,\,681\,726\,067\,169\,454\,\\
32&5\,997\,359\,460\,809\,616\,886\,494&408\,264\,709\,\,609\,407\,519\,880\,320\,\\
33&28\,233\,744\,272\,563\,685\,150\,118&1\,993\,794\,711\,\,631\,386\,183\,977\,574\,\\
34&132\,853\,629\,626\,823\,234\,210\,582&9\,724\,709\,261\,\,537\,887\,936\,102\,872\,\\
35&625\,248\,129\,452\,557\,974\,777\,990&47\,376\,158\,929\,\,939\,177\,384\,568\,598\,\\
36&2\,941\,370\,856\,334\,701\,726\,560\,670&230\,547\,785\,968\,\,352\,575\,619\,933\,376\,\\
\hline
\end{tabular}
\end{table}

\begin{table}
\caption{
\label{tab:fits}
Fitted values for the parameters in Eq.~(\ref{eq:fits}).}
\begin{tabular}{l|l|l|l}
$A_1$		& 1.1951966888	& $A_2$ 	& 1.3985089252\\
$\mu_1$		& 4.6840041570	& $\mu_2$ 	& 4.6835229879\\
$\theta_1$	& 0.1597395125	& $\theta_2$ 	& 1.3471657788\\
$c_1$		& 0.1227360755	& $c_2$ 	&-0.1161833307\\
$\Delta_1$	& 1.4315024046	& $\Delta_2$ 	& 0.2565969416\\
$k_1$		&-0.0619076482	& $k_2$		& 0.1737181819\\
$\alpha_1$	& 1.8985141134	& $\alpha_2$	& 3.4085635026\\
$\epsilon_1$	& 1.51e-13	& $\epsilon_2$	& 2.62e-12 
\end{tabular}
\end{table}


\begin{thebibliography}{99}

\bibitem{degennes}
P.-G. de Gennes, {\it Scaling Concepts in Polymer Physics}, (Cornell University Press, Ithaca, NY, 1979).

\bibitem{jensen}
I. Jensen,
{\it Enumeration of self-avoiding walks on the square lattice},
J. Phys. A: Math. Gen. {\bf 37}, 5503--5524 (2004).

\bibitem{clisby}
N. Clisby, R. Liang, and G. Slade, 
{\it Self-avoiding walk enumeration via the lace expansion},
J. Phys. A: Math. Theor. {\bf 40}, 10973--11017 (2007).

\bibitem{madras96}
N. Madras and G. Slade, {\it The Self-Avoiding Walk}, (Birkh\"{a}user, Boston, 1993).

\bibitem{orr47}
W. J. C Orr,
{\it Statistical treatment of polymer solutions at infinite dilution},
Trans. Faraday Soc. {\bf 43}, 12--27 (1947).

\bibitem{fisher59}
M. E. Fisher and M. F. Sykes,
{\it Excluded-volume problem and the Ising model of ferromagnetism},
Phys. Rev. {\bf 114}, 45--58 (1959).

\bibitem{guttmann87}
A. J. Guttmann,
{\it On the critical behaviour of self-avoiding walks},
J. Phys. A: Math. Gen. {\bf 20}, 1839--1854 (1987).

\bibitem{guttmann89}
A. J. Guttmann,
{\it On the critical behaviour of self-avoiding walks: II},
J. Phys. A: Math. Gen. {\bf 22}, 2807--2813 (1989).

\bibitem{macdonald92}
D. MacDonald, D. L. Hunter, K. Kelly, and N. Jan,
{\it Self-avoiding walks in two to five dimensions:
exact enumerations and series study},
J. Phys. A: Math. Gen. {\bf 25}, 1429--1440 (1992).

\bibitem{macdonald00}
D. MacDonald, S. Joseph, D. L. Hunter, L. L. Moseley, N. Jan,
and A. J. Guttmann,
{\it Self-avoiding walks on the simple cubic lattice},
J. Phys. A: Math. Gen. {\bf 33}, 5973--5983 (2000).


\bibitem{vanlint92}
J.H. van Lint and R.M. Wilson,
{\it A Course in Combinatorics},
(Cambridge University Press, Cambridge, UK, 1992).

\bibitem{pivot}
N. Clisby,
{\it Accurate estimate of the critical exponent 
$\nu$ for self-avoiding walks via a fast implementation of the pivot algorithm},
 Phys. Rev. Lett. {\bf 104}, 055702 (2010).

\bibitem{hsu}
H.-P. Hsu, W. Nadler, and P. Grassberger,
{\it Scaling of star polymers with 1-80 arms},
Macromolecules {\bf 37}, 4658--4663 (2004).

\end{thebibliography}
\end{document}